# Spin polarised injection in sexithienyl thin films

V. Dediu[@], M. Murgia, F. C. Matacotta, S. Barbanera*, C. Taliani

Istituto di Spettroscopia Molecolare, CNR, Via P. Gobetti, 101, 40129, Bologna, Italy
*Istituto di Elettronica dello Stato Solido ,CNR, Via Cineto Romano, 42, 00156 Roma, Italy.

## Abstract

Electroluminescence in organic semiconductors strongly depends on the relative population of excited singlet and triplet excitonic states resulting from recombination of injected carriers. In conventional organic light-emitting diodes (OLED) optical emission is usually based on fluorescence from excited singlet states[1], although phosphorescence from triplets is also investigated[2]. Governing the spin statistics of the injected carriers would open the way to amplify a chosen electronic transition increasing therefore the OLED efficiency. Direct injection of carriers with high degree of spin polarisation along a given axis should lead to preferential population of either singlet or triplet excited states, depending on the relative electrodes polarisation. Here we report a first experimental evidence of direct spin polarised (SP) injection in sexithienyl ($T_6$), a prototypical organic semiconductor, from half-metallic manganites. The spin penetration depth in $T_6$ is about 250-300 nm at room temperature. The results are discussed taking into account possible spin-flip mechanisms in organic material and interface effects.

In organic semiconductors the spin statistics of carrier recombination is focussing the interest of many researchers due to both fundamental and application aspects. Electroluminescence in organic light emitting diodes (OLED) is mainly based on the radiative decay of singlet excitons[1] which are created by the injection of electrons (*n*) and holes (*p*) from separate electrodes. The singlet/triplet occupation ratio is generally thought to be 1:3, considering a similar formation probability for one singlet and three triplet states[1], although a recent paper[3] indicates a higher formation cross-section for singlet states. Colour changing is allowed by switching from singlet to triplet decay based OLEDs[2]. Thus, governing the spin statistics in charge recombination processes would permit to improve the efficiency and tailor the optical properties of OLED devices.

The exciton spin statistics can be directly controlled by injection of carriers (*n, p*) with a given spin orientation. Such an SP injection would allow to preferentially form either singlet (antiparallel spin orientation for injected *n* and *p*) or triplet (parallel spin orientation for *n* and *p*) excitons.

SP carriers can be generated in different ways either indirectly, via photoexcitation by circularly polarised light[4] or directly, via real space injection from SP materials[5,6]. Direct injection is the most attractive and simplest method but it can be strongly affected by interface effects. The discovery of colossal magnetoresistance (CMR) in perovskite manganites opens the way to achieve 100% SP injectors[7].

Here we present an experiment on direct SP injection into sexithienyl ($T_6$) thin films from $La_{0.7}Sr_{0.3}MnO_3$ manganite electrodes with high spin polarisation. To the best of our knowledge this is the first evidence of SP injection and spin coherent transfer in organic semiconductors.

$T_6$, a $\pi$-conjugated rigid-rod oligomer, is an organic semiconductor which is currently one of the materials of choice for the development of organic based electronic[8]. Laser action from singlet emission has been recently demonstrated in an electrically injected $T_6$ based OLED[9]. Thin film mobility ranges from $10^{-2}$ to $10^{-4}$ $V^2cm^{-1}s^{-1}$ depending on morphology[10,11]. Resistance measurements in magnetic fields up to 1 T showed no intrinsic magnetoresistance (MR) on $T_6$ films.

$La_{1-x}Sr_xMnO_3$ (0.2<x<0.5) is a CMR ferromagnet with 100% SP carriers at $T<<T_c$ (Curie temperature $T_c$ < 370 K). The carrier hopping between $Mn^{3+}$ and $Mn^{4+}$ ions is governed by a strong on site Hund interaction (nearly 1 eV). $La_{1-x}Sr_xMnO_3$ easily accepts carriers with a spin parallel to the average spin polarisation and offers a colossal resistance to the carriers with opposite spin polarisation[12]. $La_{1-x}Sr_xMnO_3$ is a peculiar metal[13] characterised by a narrow band (<1 eV) and a small carrier density ( about $10^{21}$ $cm^{-3}$) like in high temperature cuprate superconductors and organic metals.

A spin-valve experiment was performed on hybrid structures shown in Fig. 1. $La_{0.7}Sr_{0.3}MnO_3$ epitaxial thin films were deposited on suitable insulating substrates by using the Channel-Spark deposition method[14]. Simple planar structures with an electrode separation ranging from 70 to 500 nm were fabricated by electron-beam lithography. $T_6$ thin films (100-150 nm thick) were deposited by molecular beam deposition[15]. In the absence of the external magnetic field the electrodes have random spin orientation with respect to each other. Applying a magnetic field the spins in both electrodes orient parallelly. For a direct contact between FM electrodes this results in a strong negative MR. A negative MR across the organic semiconductor would indicate both a SP injection into organic and a SP coherent transport between the electrodes.

Figure 2 shows the magnetic field dependence of the I-V characteristics of a $La_{0.7}Sr_{0.3}MnO_3/T_6/La_{0.7}Sr_{0.3}MnO_3$ junction measured in ambient atmosphere and room temperature. All the I-V curves were nearly ohmic, indicating a negligible interface barrier for carrier injection. Changing the mutual orientation of SP electrodes from random to parallel by applying a magnetic field (3.4 kOe), a strong decrease of the device resistance was observed. The negative MR signal is nearly 30% for a channel length of 140 nm. Such a negative MR is unexpected for the bare $T_6$ material (see above), and indicates that spin polarisation is maintained inside the organic semiconductor. By removing and re-depositing the organic film with suitable solvents a good reproducibility was found (Fig.2). The observed MR response is comparable, if not higher, than in commercial spintronic devices[16]. A higher MR may be achieved by improving the spin alignment contrast between the electrodes. By increasing the gap the MR decreases to 7-10% for a 200-nm gap and vanishes at about 300 nm. The room temperature spin penetration depth is therefore estimated to be about 200-300 nm. Electrical resistivity of $10^6$ Ohm cm indicates that the $T_6$ films are indeed weakly doped by ambient oxygen[17]. The spin relaxation time estimated for an $10^{-4}$ $V^2cm^{-1}s^{-1}$ mobility is about 1 μs.

The observed SP effect can be divided in two separate processes: SP injection into organic material and coherent spin transport across it.

The SP injection into inorganic semiconductors has been extensively studied both theoretically and experimentally. In the case of FM metal/(inorganic) semiconductor boundary it was found that only nearly 1% of the injected carriers conserve their spin orientation[18]. This can be understood taking into account that the highest resistance in the system (semiconductor resistance $R_S$) is spin independent, so it is difficult to maintain an SP current[19]. To avoid this problem ferromagnetic semiconductors were used as spin injecting materials[20]. This significantly increases the SP current but requires low temperatures, due to low $T_c$. It has been recently proposed that the insertion of a spin selective interface resistance between the FM metal and the semiconductor may solve the problem[21,22]. This resistance $R_{int}$ should be higher than both $R_{FM}$ and $R_S$. The CMR/$T_6$ interface may give rise to a tunnelling barrier as it the case in various organic/metal interfaces[17], while the spin selectivity of such interface is due to boundary with the CMR material.

As far as spin coherent transport in organic materials is concerned we shall analyse the possible spin-flip mechanisms. The most important ones are spin-orbit interaction and hyperfine interaction. In a π-conjugated oligomer only *p*-orbitals are delocalised. The wavefunctions for these orbitals have a zero amplitude on nucleus site, minimising the effect of hyperfine interaction. A sizeable spin-orbit coupling is provided by sulphur atoms but these atoms seem to play a marginal role in the carrier transport[23].

In conclusion, the evidence of SP injection and polarised spin transport through organic semiconductors (up to 200-300 nm) has been demonstrated for the first time. Moreover the ambient atmosphere and room temperature operation of this process opens the possibility of practical applications. The benefit of spin polarised injection is expected to have a strong impact on organic optoelectronics by governing the singlet-triplet exciton ratio. On the other hand this opens the way for the use of organic materials in such a new and important field as Spintronics.

@e-mail: dediualek@hotmail.com

References

1. Friend, R. H. *et al.* Electroluminescence in conjugated polymers. *Nature* **397**, 121-128 (1999).
2. Baldo, M. A., Lamansky, S., Burrows, P. E., Thompson, M. E. & Forrest, S. R. Very high-efficiency green organic light-emitting devices based on electrophosphorescence. *Appl. Phys. Lett.* **75**, 4-6 (1999).
3. Wohlgenannt, M., Tandon, K., Mazumdar, S., Ramasesha, S. & Vardeny, Z. V. Formation cross-section of singlet and triplet excitons in π-conjugated polymers. *Nature* **409**, 494-497 (2001).
4. Hägele, D., Oestreich, M., Rühle, W. W., Nestle, N. & Eberl, K. Spin transport in a GaAs. *Appl. Phys. Lett.* **73**, 1580 (1998).
5. Johnson, M. & Silsbee, R. H. Spin-injection experiment. *Phys. Rev. B.* **37**, 5326-5335 (1988).
6. Ohno, Y. *et al.* Electrical spin injection in a ferromagnetic semiconductor heterostructure. *Nature* **402**, 790-792 (1999).
7. Park, J.-H. *et al.* Direct evidence for a half-metallic ferromagnet. *Nature*, 794-796 (1998).
8. Dodabalapur, A., Torsi, L. & Katz, H. E. Organic transistors: two-dimensional transport and improved electrical characteristics. *Science* **268**, 270-271 (1995).
9. Schoen, J. H., Kloc, C., Dodabalapur, A. & Batlogg, B. A light-emitting field effect transistor. *Science* **290**, 963-965 (2000).
10. Torsi, L., Dodabalapur, A., Rothberg, L. J., Fung, A. P. & Katz, H. E. Charge transport in oligothiophene field-effect transistors. *Phys. Rev. B* **57**, 2271-2275 (1998).
11. Ostoja, P. *et al.* Instability in electrical performance of organic semiconductor devices. *Adv. Mater. Opt. Electr.* **1**, 127-132 (1992).
12. de Gennes, P. G. Effects of double exchange in magnetic crystals. *Phys. Rev. B* **118**, 141-154 (1960).
13. Rao, C. N. R. *Colossal Magnetoreistance, Charge Ordering and Related Properties of Manganese Oxides* (eds. Rao, C. N. R. & Raveau, B.) (World Scientific, Singapore, 1998).
14. Dediu, V. *et al.* Micro-Raman and Resistance Measurements of Epitaxial $La_{0.7}Sr_{0.3}MnO_3$ Films. *Phys. Stat. Sol. (b)* **215**, 625-629 (1999).
15. Murgia, M. *et al.* In-Situ Characterisation of the Oxygen Induced Changes in a UHV Grown Organic Light-Emitting-Diode. *Synth. Met.* **102**, 1095 (1999).
16. Prinz, G. A. Magnetoelectronics. *Science* **282**, 1660-1663 (1998).
17. Fichou, D. & Ziegler, C. in *Handbook of Oligo- and Polythiophenes* (ed. Fichou, D.) (Wiley-VCH, Weinheim, 1998).
18. Hammar, P. R., Bennett, B. R., Yang, M. J. & Johnson, M. Observation of Spin Injection at a Ferromagnet-Semiconductor Interface. *Phys. Rev. Lett.* **83**, 203-206 (1999).
19. Schmidt, G., Ferrand, D., Molenkamp, L. W., Filip, A. T. & van Wees, B. J. Fundamental obstacle for electrical spin injection from a ferromagnetic metal into a diffusive semiconductor. *Phys. Rev. B* **62**, 4790-4793 (2000).
20. Fiederling, R. *et al.* Injection and detection of a spin-polarized current in a light-emitting diode. *Nature* **402**, 787-790 (1999).
21. Rashba, E. I. Theory of electrical spin injection:Tunnel contacts as a solution of the conductivity mismatch problem. *Physical Review B* **62**, R16267-R16270 (2000).
22. Kirczenow, G. Ideal spin filters: A theoretical study of electron transmission through ordered and disordered interfaces between ferromagnetic metals and semiconductors. *Phys. Rev. B* **63** (2001).
23. Bennati, M., Nemeth, K., Surjan, P. R. & Mehring, M. Zero-field-splitting and π–electron spin densities in the lowest excited triplet state of oligothiophenes. *J. Chem. Phys.* **105**, 4441-4447 (1996).

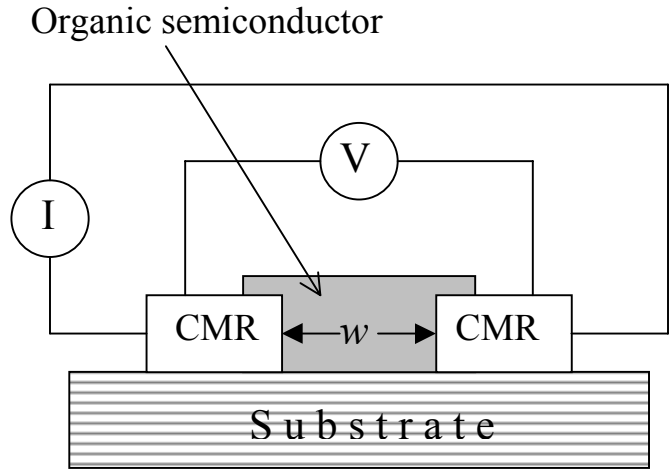

Fig. 1. A schematic structure of a planar CMR/T$_6$/CMR device.

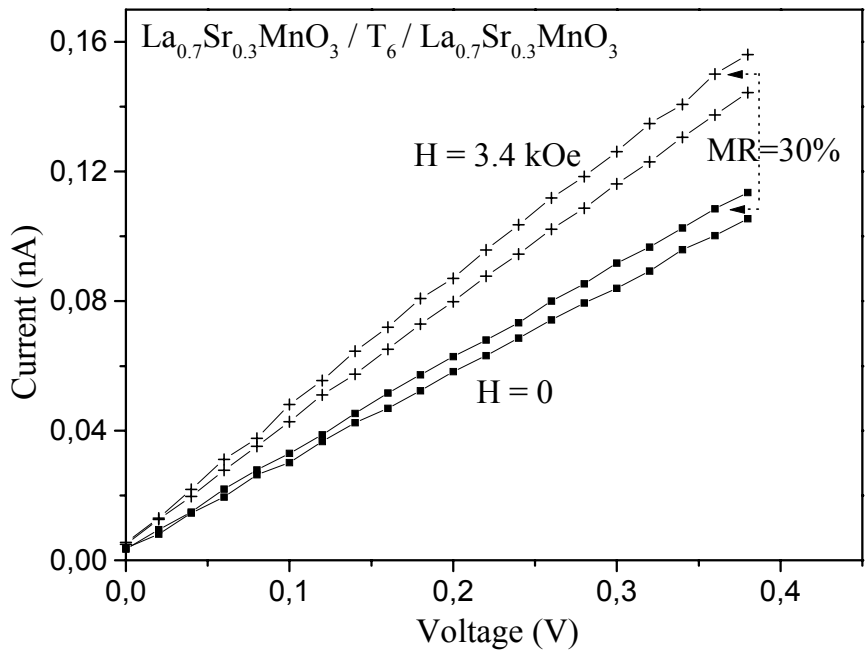

Fig. 2. I-V characteristics of a La$_{0.7}$Sr$_{0.3}$MnO$_3$/T$_6$/ La$_{0.7}$Sr$_{0.3}$MnO$_3$ junction as a function of magnetic field at ambient atmosphere and room temperature. The La$_{0.7}$Sr$_{0.3}$MnO$_3$ thin film is 100 nm thick and the gap between the electrodes is 140 nm. The T$_6$ overlayer is 150 nm thick. Squares indicate measurements in the absence of magnetic field, crosses indicate measurements at 3.4 kOe. Different curves indicate extreme deviations for different organic films on the same gap obtained by subsequent dissolution and re-deposition of the organic thin film.